\documentclass[aps,pra,preprint,groupedaddress,showpacs]{revtex4}
\usepackage{graphicx}
\usepackage{dcolumn}
\usepackage{bm}

\begin{document}

\title{Atom interferometry using wavepackets with constant spatial displacements}

\author{Edward J. Su$^{1}$}
\altaffiliation[current address: ]{Dept. Phys., MIT, Cambridge, MA, 02139}

\author{Saijun Wu$^{1,2}$}
\altaffiliation[current address: ]{NIST, Gaithersburg, MD, 20899}

\author{Mara G. Prentiss$^{1}$}

\affiliation{$^1$Harvard University Department of Physics, Cambridge
Massachusetts 02138}

\affiliation {$^2$Harvard University School of Engineering and
Applied Science, Cambridge Massachusetts 02138}
\date{\today}

\begin{abstract}

We demonstrate a standing-wave light pulse sequence that places atoms into a superposition of wavepackets with precisely controlled
displacements that remain constant for times as long as 1 s. The separated wavepackets are subsequently recombined resulting in atom
interference patterns that probe energy differences of $\approx 10^{-34}$~Joule, and can provide acceleration measurements that are insensitive
to platform vibrations.

\end{abstract}
\pacs{39.20+q, 03.75.Dg} \maketitle
\section{Introduction}

Atom interferometry employs the interference of atomic de Broglie waves for precision measurements~\cite{AIBerman}. In practice two effects
limit the ultimate sensitivity of devices where the interfering atomic wavepackets are allowed to propagate in free space: the effect of
external gravitational fields upon the atomic trajectories, and transverse expansion of the atom cloud. By accounting for gravity, atomic
fountains can increase the interrogation time during which the interferometry phase shifts accumulate~\cite{Chugravity}; alternatively one can
use magnetic dipole forces to balance the force of gravity~\cite{Clauser88}. Magnetic waveguides~\cite{CornellGuide99, PrentissGuide2000} can
trap atoms for times longer than a second, suggesting the possibility of measuring energy differences between interfering wavepackets with an
uncertainty $< \hbar$/(1 s)$\sim 10^{-34}$~Joule; however, this remarkable precision cannot be obtained if the decoherence time of the atoms is
much shorter than the trap lifetime. Early atom interferometry experiments using atoms confined in magnetic waveguides showed that the external
state coherence of the atoms decayed quite quickly, limiting interferometric measurements to times
$<$10~ms~\cite{WangMichelson05,GuideTalbot05}. More recent experiments using Bose condensates~\cite{ChipLongCoherence} have shown that the
external state coherence can be preserved for approximately 200~ms, where the decoherence is dominated by atom-atom interactions. Interferometry
experiments using either condensed atoms in a weak trap, or using non-condensate atoms in a waveguide with precise angular alignments have been
shown to have phase-stable interrogation time of $\approx 50$~ms, where the dephasing is induced by inhomogeneities in the confining
potential.~\cite{Garcia06,Burke08,Burke09,movingguide}.

This work demonstrates a new atom interferometer configuration that measures the differential phase shift of spatially-displaced wavepacket
pairs. We demonstrate phase-stable interferometry operations with up to one second interrogation time by applying the technique to atoms in a
straight magnetic guide. We show that the matter-wave dephasing rate scales linearly with the wavepacket displacement, suggesting that dephasing
in our interferometer is primarily caused by a weak longitudinal confinement of the atoms. We also demonstrate that the phase readout of the
interferometer is less sensitive to vibration than conventional interferometery schemes, which should enable precision measurements even in
noisy environments such as moving platforms.

\begin{figure}
\centering
\includegraphics [width=3 in,angle=0] {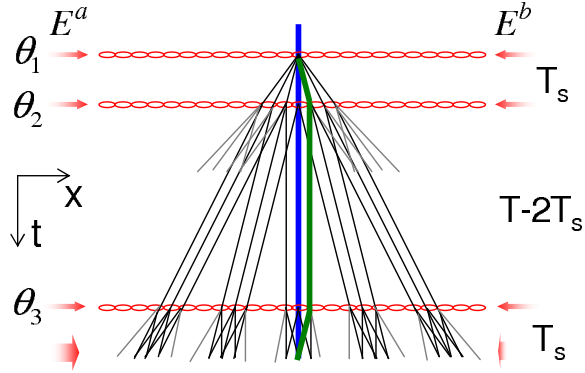}\\
\caption{(Color online) Recoil diagram of the 4-pulse scheme. The light shift potential of the standing wave is represented by the array of circles. Straight
lines show the centers of the diffracted wavepackets. A representative pair of interfering paths is marked with thick lines. The matter-wave
interference at around $t=T$ is probed with $E^a$. The backscattered light, e.g., the ``grating echo'' $E^b(\tau)$, is mixed with a weak local
field from $E^b$ mode for phase retrieval.}\label{figdiagram}
\end{figure}

\section{The 4-pulse grating echo scheme}

Typical Talbot-Lau matter-wave interferometery~\cite{ClauserInftheory92,TLCahn97, Periodic02,C70Inf} employs a 3-grating diffraction scheme. In
the most common time-domain setup, an atomic wavepacket is diffracted by a periodic potential, applied briefly at time $t=0$, into a collection
of wavepackets that depart from each other at multiples of the velocity ${\bf v}_{\bf Q}=\hbar {\bf Q}/m$, where ${\bf Q}$ is the potential's
wavevector and $m$ is the atomic mass. The potential is pulsed on again at $t=T/2$, and the different velocity classes created by the first
pulse, which have now moved away from each other, are each again diffracted into multiple orders. After the second pulse there will be pairs of
wavepackets whose relative velocity has been reversed from before; these will move toward each other, then overlap and interfere near time
$t=T$. Those with relative velocity $n{\bf v}_{\bf Q}$ will generate density fringes with wavevector $\pm n{\bf Q}$.

As in earlier experiments~\cite{TLCahn97,gratingechoOrigin} we use an off-resonant optical standing wave (SW) to create the pulsed periodic
potentials, and observe the resulting atomic density fringes by measuring the Bragg scattering of an optical probe. In our experiment we observe
the lowest order fringes, $n=\pm 1$, where the Bragg condition corresponds approximately to backscattering of one of the beams that forms the
standing wave.

The interferometric technique presented here employs a 4-pulse scheme (Fig.~\ref{figdiagram}), where the additional pulse is used to halt the
relative motion of the interfering wavepacket pairs. After the first pulse is applied the situation is identical to that in the 3-pulse case,
with the original wavepacket split into different diffraction orders corresponding to velocities $\pm n{\bf v}_{\bf Q}$. Here we quickly apply a
second pulse after a short time $t=T_s$. The pairs of wavepackets that we are interested in, those that will eventually interfere, are those
that now have zero momentum difference; these pairs have the same velocity but are displaced in space, having moved apart by a distance (${\bf
d} = {\bf v_Q} T_s$) during the interval between pulses. After waiting for a time $T-2T_S$ the coherence between the wavepackets is measured by
allowing them to interfere; a third pulse diffracts the wavepackets at $t=T-T_s$ and the resulting interference fringe is probed around $t=T$.
This 4-pulse sequence can be imagined as a 3-pulse sequence of length $2T_s$ that is "paused" between times $T=T_S$ and $T-T_S$; though during
this time the relative phase of each separated wavepacket pairs will continue to be sensitive to external fields.

The existing theory of Talbot-Lau interferometry can be straightforwardly extended to this scheme. We consider the SW field formed by the
traveling light fields $E^{a}$ and $E^{b}$ with associated k-vectors ${\bf k}^{a}$ and ${\bf k}^{b}$, with ${\bf Q}={\bf k}^a-{\bf k}^b=Q {\bf
e_x}$. Experimentally, we measure the backscattering of light from ${\bf k}^{a}$ into ${\bf k}^{b}$; this is characterized by an electric field
component $E^b(\tau)$ that can be expressed in terms of the atomic density operator:
\begin{equation}
E^b (\tau) =  - ig E_p {\rm Tr}[\hat \rho(T) e^{i Q (\hat x+ \frac
{\hat p \tau}{m})}]. \label{backamp}
\end{equation}
Here $E_p$ gives the amplitude of probe light from the $E^a$ mode, $g$ is a constant that depends on the atomic polarizability and the number of
atoms participating in the interaction, $\hat x$ and $\hat p$ are the position and momentum operators for atomic motion along $\bf e_x$, and
$\hat \rho(T)$ is the single atom density matrix at time $T$. Consider an atomic sample with a rms velocity $u$ and a thermal de Broglie
wavelength $l_c=\hbar/(m u)$, in Eq.~(\ref{backamp}) we use $\tau \sim l_c/v_Q$ to specify a coherence-length-dependent time window $(T-l_c/v_Q,
T+l_c/v_Q)$ during which the atomic wavepackets overlap so that the interference fringe contrast is non-zero. Experimentally, the amplitude of
$E^b(\tau)$ is averaged during this time window to extract the magnitude of the interference fringe; this will be referred to as the amplitude
of the ``grating echo'', $E_g(T_s,T):=<E^b(\tau)>_{\tau}$~\cite{swthesis}.

For a SW pulse in the Raman-Nath regime, where the atomic motion can be neglected for its duration, the $n^{th}$ order matter-wave diffraction
is weighted by the amplitude $i^n J_n(\theta)$, with $J_n$ the $n^{th}$ order Bessel function and $\theta$ the time-integrated light shift or
pulse area. We define $\{T_1,T_2,T_3, T_4\}=\{0,T_s,T-T_s, T\}$, and specify the position of standing wave nodes at $T_i$ with the SW phase
$\varphi_i$. The interaction of the first three SW pulses in the 4-pulse interferometer can be effectively described by
\begin{equation}
\hat V_{\rm SW}(t)=\hbar \sum\nolimits_{i=1}^{3}\theta_i
\delta(t-T_i) \cos(Q \hat x+\varphi_i). \label{equsw}
\end{equation}
Since the standing wave phases $\varphi_i$ involve simple algebra, we will ignore them during the following discussion, and
reintroduce them when they become relevant.

In order to account for imperfections in the guide we consider the 1D motion of atoms along $\bf e_x$ during the interferometry sequence to be
governed by $\hat H= \hat H_0 +\hat V_{\rm SW}(t)$ and further $\hat H_0=\hat p^2/2 m + V(\hat x)$ where $V(\hat x)$ is a general 1D potential.
We introduce the time-dependent position and momentum operators $\hat x(t)=e^{i \hat H_0 t} \hat x e^{-i \hat H_0 t}$ and $\hat p(t)=e^{i \hat
H_0 t} \hat p e^{-i \hat H_0 t}$. For $T_s$ to be sufficiently short, the atomic motion can be treated as free during $0<t<T-T_s$ and
$T-T_s<t<T$. For a thermal atomic sample with $Q l_c<<1$, we find at the leading order of $Q l_c$, the interferometer output is related to the
initial atomic density matrix $\hat \rho(0)$ by
\begin{equation}
\begin{array}{l}
E_g(T_s,T) =- ig E_p J_{ - 1}( 2\theta _3 \sin \omega_Q T_s) \times \\
\sum\limits_{m_1 ,m_2 } c_{m} {\rm Tr}[\hat\rho(\delta
x_m,\delta p_m,0) e^{\frac{i}{\hbar} d \hat p(T)} e^{-\frac{i}{\hbar} d \hat p(0)}].\\
\end{array}\label{backamp2}
\end{equation}
Here $c_{m} = J_{m_1 }(\theta_1) J_{m_1  + 1} (\theta_1)J_{m_2
}(\theta_2) J_{m_2  - 1}(\theta_2) e^{i\phi_m}$, $\phi_m= (2 m_1+1)
Q l_c + (2 m_2-1) \omega_Q T_s$, $\delta x_m=m_1 v_Q T_s$, $\delta
p_m=(m_1+m_2)Q$. In Eq.~(\ref{backamp2}) $\omega_Q$ is the
two-photon recoil frequency of atoms and we have $d=v_Q T_s$.

The second line of Eq.~(\ref{backamp2}) composes a weighted sum of matter-wave correlation functions. The initial conditions of matter-wave
states are specified by a density matrix $\hat\rho(\delta x,\delta p,0)=e^{i(\delta x \hat p+\delta p \hat x)} \hat \rho e^{-i(\delta x \hat
p+\delta p \hat x)}$ that describes an atomic ensemble that is identical to $\hat\rho(0)$, but with mean position and momentum shifted by
$\delta x$ and $\delta p$ respectively. The correlation function gives the average overlap of wavepacket pairs propagating under an external
potential displaced by $d \bf e_x$, with one example sketched with the thick lines in Fig.~\ref{figdiagram}. The correlation functions are in
direct analogy to the neutron scattering correlation function discussed in ref.~\cite{phasespaceDispDiego} where momentum displacements were
considered.  Notice that if the uniform atomic sample has a spatial extension $L>>\delta x$ and with thermal velocity $u>>\delta p/m$, the
original and shifted density matrix are approximately the same, and the correlation functions are approximately independent of $\delta x_m$ or
$\delta p_m$. We can thus use a sum rule of Bessel functions to simplify the second line of Eq.~(\ref{backamp2}) giving:

\begin{equation}
\begin{array}{l}
E_g(T_s,T)\approx  - i \tilde g J_{ - 1}^2 (2\theta _2 \sin \omega_Q
T_s )\times \\{\rm Tr}[\hat \rho (0)e^{\frac{i}{\hbar} d \hat p(T)}
e^{ - \frac{i}{\hbar}d \hat p(0)}],
\end{array}\label{backcompact}
\end{equation}
where we have chosen $\theta_2=\theta_3$ and define $\tilde g=g E_p
\theta _1 Q l_c$.

Three features of Eq.~(\ref{backamp2}) and Eq.~(\ref{backcompact})
are worth noting:

First, though we have only considered the 1D motion of atoms in the
external potential $V(x)$, the formula is readily applicable to a 3D
time-dependent potential $V({\bf r}, t)$ as long as the external
potential contributes negligibly to the differential phase shift of
wavepacket pairs during $0<t<T_s$ and $T-T_s<t<T$.

Second, the reduction from Eq.~(\ref{backamp2}) to
Eq.~(\ref{backcompact}) requires that the correlation functions be
insensitive to momentum transferred by the SW pulses. This is very
well satisfied if the displacements are much smaller than the
position and momentum spreadings of the atomic gas itself since
$\hat\rho(\delta x,\delta p,0)\approx \hat \rho(0)$. In addition,
the approximation is particularly well satisfied if the potential is
periodic at small wavelengths~\cite{kickedrotor95,stability08}.

Finally, notice that both Eq.~(\ref{backamp2}) and
Eq.~(\ref{backcompact}) can be evaluated semi-classically by
replacing $ {\rm Tr}[\hat \rho (0)e^{\frac{i}{\hbar} d \hat p(T)}
e^{ - \frac{i }{\hbar}d \hat p(0)}]$ with $<e^{\frac{i m d}{\hbar}
[v(T)- v(0)]}>_c$, where $v(t)$ gives the classical velocity of
atoms along $\bf e_x$ and $<...>_c$ gives the classical ensemble
average over the atomic initial conditions.

Now consider a weak quadratic potential $V(x)=m a x+\frac{1}{2}m\omega_l^2 x^2$ with an acceleration force $m a$ and with $\omega_l T<<1$ to
model the potential variation along the nearly free (axial) direction of propagation in the magnetic guide, also assuming an atomic sample with
a Gaussian spatial distribution along $\bf e_x$ given by $\rho(x)=e^{-x^2/(2L^2)}$. The expected amplitude of the grating echo signal is then
found to oscillate with $T_s$ and to decay as a Gaussian with the total interrogation time $T$:
\begin{equation}
\begin{array}{l}
E_g(T_s,T)\approx -i \tilde g J_{ - 1}^2 (2\theta _2 \sin \omega_Q
T_s )\times \\e^{-\frac{1}{2}(\frac{m d}{\hbar}\omega_l^2 T L)^2}
e^{i (\frac{m d }{\hbar}a T+
\varphi_{1,2}-\varphi_{3,4})}.\end{array}\label{backampexp}
\end{equation}
(Here we have reintroduced the standing wave phase in Eq.~(\ref{equsw}), where $\varphi_{i,j}=\varphi_i-\varphi_j$)

\begin{figure}
\centering
\includegraphics [width=3.2 in,angle=0] {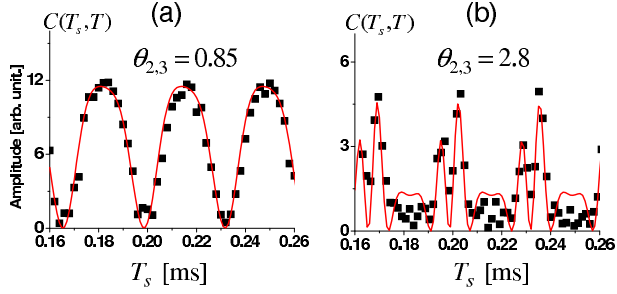}\\
\caption{(Color online) Examples of interferometry signal amplitude $C(T_s, T)$
oscillation vs $T_s$ at fixed $T$. Scatter plots are from
experiments. Solid lines are calculated according to an extension of
Eq.~(\ref{backampexp}) with complex pulse area $\tilde
\theta=\theta(1+0.025i)$. (a): $\theta=0.85$ and $T=50$~ms. (b):
$\theta=2.8$ and $T=30$~ms. The data were taken during different
period of experiments with different signal/noise. }\label{figTso}
\end{figure}

\section{Experimental setup}

The experimental apparatus is described in detail in Ref. ~\cite{swthesis}. A straight 2D quadruple magnetic field with a transverse gradient of
70~G/cm is generated by four 200~mm$\times$100~mm$\times$1.56~mm permalloy foils poled in alternating directions. Approximately $10^{7}$
laser-cooled $^{87}$Rb atoms in the ground state F=1 hyperfine level are loaded into this magnetic guide, resulting in a cylindrically-shaped
atom sample 1~cm long and 170~$\rm \mu$m wide. The average transverse oscillation frequency of the atoms in the guide is on the order of 80 Hz,
estimated by displacement induced oscillations of the atomic sample using absorption images. A very weak harmonic potential along the guiding
direction is estimated to be $\omega_l\sim 2\pi\times0.08$~Hz~\cite{swthesis}.

The SW fields formed by two counter-propagating laser beams with diameters of 1.6~mm are aligned to form a standing wave with k-vector along the
magnetic guide direction $\bf e_x$. Precise angular adjustment is achieved by tuning the orientation of the magnetic guide using two rotation
stages to within $2\times 10^{-4}$ radians. The optical fields are detuned 120 MHz above the $F=1$ - $F'=2$ D2 transition. We choose the SW
pulse with typical pulse area of $\theta\sim 0.8 - 3.0$, and with duration of 300~ns to be deep in the thin-lens regime of the 25~$\mu$K atomic
sample. With this pulse duration, the fraction of atoms contributing to the final interference fringe is typically limited by SW diffraction
efficiency to about ten percent. We probe the $\lambda/2$ atomic density grating at around time $t=T$ by turning on only one of the traveling
wave beams; the other beam is attenuated and shifted by 6~MHz to serve as an optical local oscillator, where the combined intensity is measured
using a fiber-coupled avalanche photodetector. The beat signal is measured and numerically demodulated using the 6~MHz rf reference to recover
the grating echo signal $E_g(T_s,T)=C(T_s,T)e^{i\varphi(T_s,T)}$. The interferometer signal amplitude $C(T_s,T)$ and phase $\varphi(T_s,T)$ are
measured for different interferometer parameters.

\section{Results and discussions}

According to Eq.~(\ref{backampexp}), the pre-factor $J_1^2(2\theta\sin\omega_Q T_s)$ in the backscattering amplitude is an oscillatory function
of $T_s$, with the periodicity determined by $2\omega_Q=2\pi/33.15$~$\rm \mu$s$^{-1}$. The amplitude oscillation is reproduced experimentally;
two examples are plotted in Fig.~\ref{figTso}, where $T_s$ is varied from 0.16~ms to 0.26~ms. In Fig.~\ref{figTso}(a) a relatively small SW
pulse area $\theta\approx 0.85$ was chosen so that the Bessel function is approximately linear. Correspondingly, we see the oscillation is
approximately sinusoidal. In Fig.~\ref{figTso} (b) a strong SW pulse with area $\theta\approx 2.8$ was chosen and the Bessel function becomes
highly nonlinear. Nevertheless, the experimental data still fits the theoretical expectation from Eq.~(\ref{backampexp}) fairly well. The values
for the of pulse area in the calculation were found in agreement with the SW pulse intensity-duration products. Notice the solid lines in
Fig.~\ref{figTso} were calculated according to an extension of Eq.~(\ref{backampexp}) with a complex SW pulse area including an imaginary part
to account for the optical pumping effect at the 120~MHz SW detuning~\cite{swthesis}.

\begin{figure}
\centering
\includegraphics [width=3.3 in,angle=0] {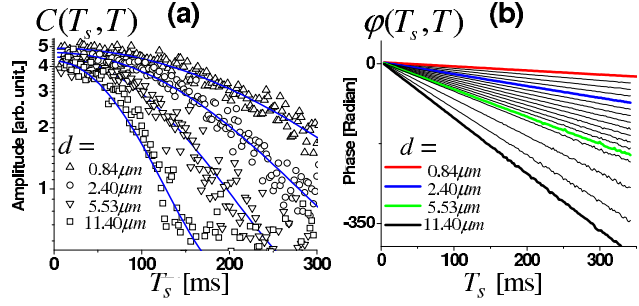}\\
\caption{(Color online) (a): Backscattering amplitude decay. The scatter plots give
$C(T)$ at four different displacement $d$. The solid lines are due
to Gaussian fit. (b): Interferometry phase shift $\varphi(T)$ vs $T$
at different displacement $d$. Four of the data traces from the same
experiments as those in (a) are plotted with thick
lines.}\label{figampvsphase}
\end{figure}

With fixed $T_s$ at the peak values of the amplitude oscillations, we now consider the dependence of the interferometer signals on the total
interrogation time $T$. Fig.~\ref{figampvsphase} gives examples of the interferometer amplitude decay $C(T)$ and phase shift $\varphi(T)$ at
various $d= v_Q T_s$. From Fig.~\ref{figampvsphase}(a) we see the amplitude decay is slower for smaller $d$, while all the $C(T)$ fit fairly
well to Gaussian decay, in agreement with Eq.~(\ref{backampexp}) derived from a weak harmonic confinement model. In Fig.~\ref{figampvsphase}(b)
we see the phase readout is a linear function of interrogation time $T$, also in agreement with Eq.~(\ref{backampexp}). By applying
Eq.~(\ref{backampexp}) to the observed phase shifts, we consistently retrieve an acceleration $a = 83.4$~mm/s$^2$ for different $d$. The
acceleration is due to a small component of gravity along the standing-wave/magnetic guide direction $\bf e_x$~\cite{swthesis}, as confirmed by
varying the tilt angle of the apparatus.

\begin{figure}
\centering
\includegraphics [width=3.3in,angle=0] {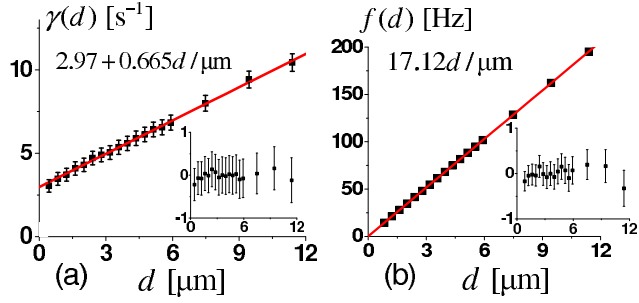}\\
\caption{(Color online) (a): Backscattering amplitude decay rate $\gamma(d)$
[extracted from a Gaussian fit of $C(T)$ data such as those from
Fig.~\ref{figampvsphase} (a)] vs wavepacket displacement $d$. (b):
Interferometry phase shift rate $f(d)=\frac{d \varphi(T)}{d T}$
[extracted from a linear fit of $\varphi(T)$ data from
Fig.~\ref{figampvsphase} (b)] vs $d$. The insets give the residuals
of the linear fits.}\label{figfit}
\end{figure}

\begin{figure}
\centering
\includegraphics [width=3.2in,angle=0] {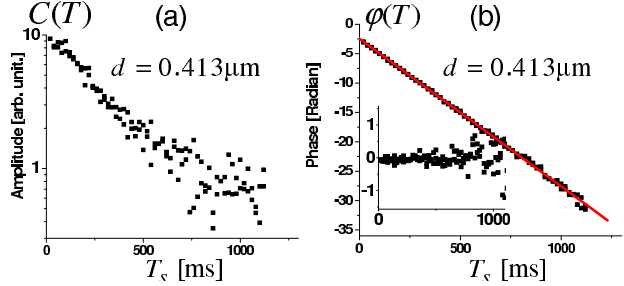}\\
\caption{(Color online) Interferometry readouts $C(T)$ and $\varphi(T)$ at
$d=0.413$~$\mu$m. (a): $C(T)$ vs $T$; (b): $\varphi(T)$ vs $T$, with
inset giving the residual after a linear fit.}\label{figonesec}
\end{figure}

We extract the amplitude decay rate $\gamma(d)$ by fitting the $C(T)$ decay data with $C(T)\propto e^{-(\gamma(d)T)^2}$. The dephasing rate
$\gamma(d)$ is plotted vs the displacement $d$ in Fig.~\ref{figfit}(a). The  $d$-dependence of $\gamma(d)$ shows good agreement with a linear
fit. According to Eq.~(\ref{backampexp}), for weak confinement along $\bf e_x$ with $\omega_l\sim 2\pi\times 0.08$~Hz and for $L\sim 2.5$~mm of
our 1~cm atomic sample, we expect $\gamma(d)\sim m \omega_l^2 d L/\hbar/\sqrt{2}\sim 0.6 d/$~$\mu$m~s$^{-1}$. This agrees with the
experimentally measured $\gamma (d) = (2.97 + 0.665d/\mu$m)~s$^{ - 1}$ according to Fig.~\ref{figfit}(a). The offset of $\gamma(d\rightarrow
0)=2.97$s$^{-1}$ is partly due to the escape of atoms from the interaction zone via collisions with the walls of the 4~cm vacuum glass cell,
which, if fit to a Gaussian, would give $\tilde \gamma(d\rightarrow0)\sim 1.6$~s$^{-1}$. The remaining discrepancy is likely due to the
inaccuracy of the Gaussian fit which is based on the assumption of a weak harmonic perturbation $V(x)$ in Eq.~(\ref{backampexp}). For small $d$
and thus a small dephasing rate, local anharmonicity in $V(x)$ might become important. Indeed, for long interaction time $T$ the decay exhibits
an exponential feature, which is clearly seen in Fig.~\ref{figonesec}(a) where the amplitude decay $C(T)$ with $d=0.418$~$\rm \mu$m and
$T_s=35.4$~$\rm \mu$s is plotted. For such a small wavepacket displacement $d$, the phase of the backscattering signal remains stable for
$T>1$~s, as shown in Fig.~\ref{figonesec}(b).

\begin{figure}
\centering
\includegraphics [width=3.2in,angle=0] {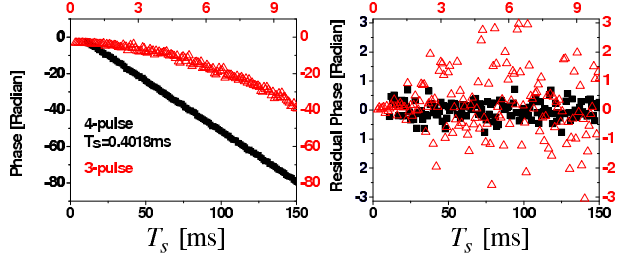}\\
\caption{(Color online) Interferometry phase readouts $\varphi(T)$ for a 3-pulse
(open triangles, right and top axes.) and a 4-pulse (solid squares,
left and bottom axes. $T_s=0.4018$~ms.) configurations. Fig. (a)
gives the phase readouts. Fig. (b) gives the residual of a quadratic
(3-pulse) and a linear (4-pulse) fit.}\label{figvib}
\end{figure}

Last, we consider the effect of phase noise in the SW on the sensitivity of our device, induced for example by vibrations of mirrors in the SW
path. For $T>>T_s$, the standing wave phase variation due to time dependent changes in mirror positions is given by $\varphi _{1,2}$, which is
not correlated with $\varphi _{3,4}$. If we specify the SW phase at time $t$ with $\phi(t)$ such that $\varphi_i=\phi(T_i)$, the mirror
vibration induced interferometer phase noise is given by $ N_{\varphi, {\rm mirror}} (T_s ,T ) = \sqrt {2 < [\phi (t) - \phi (t + T_s )]^2
> _t }$, which does not depend on T. This is different from a 3-pulse atom interferometer with mirror-induced phase noise given by
$N^{\prime}_{\varphi ,{\rm mirror}} (T) = \sqrt { < [\phi (t) - 2\phi (t + T/2) + \phi (t + T)]^2 > _t }$, where increases in
sensitivity due to increases in interaction time necessarily also result in increases in phase noise.  In contrast, in the four pulse
scheme considered here $T$ can be increased to improve the sensitivity, while keeping $N_{\varphi, {\rm mirror}} (T_s )$ unaffected.

This is illustrated in Fig.~\ref{figvib} where we compare the 3-pulse and the 4-pulse interferometer phase readouts under the same noisy
environmental conditions. A white noise voltage source is filtered to eliminate frequencies below 100~Hz, then amplified and applied to a
piezo-driven mirror in the SW optical path. As shown in Fig.~\ref{figvib}, the mirror vibration randomizes the phase of the 3-pulse
interferometer for T greater than 5~ms. Under the same conditions, the phase of the 4-pulse interferometer is stable for times longer than
150~ms. In this case the acceleration sensitivity of the 4-pulse interferometer $\delta \varphi/\delta a\sim$1~rad/mm/s$^2$ at $T=150$~ms,
exceeds that for the 3-pulse case of $\delta \varphi/\delta a\sim$0.4~rad/mm/s$^2$ at $T=10$~ms. The insensitivity of the 4-pulse scheme to
low-frequency mirror vibrations is a feature of speed-meters, as shown in Eqs.~(\ref{backcompact}),~(\ref{backampexp}) in the semiclassical
limit with the phase proportional to the velocity  during the interrogation time $T$.

\section{Summary}

We have demonstrated a 4-pulse grating echo interferometer scheme to study the dephasing effects for atoms confined in a magnetic guide. We find
linearly reduced dephasing rate at reduced wavepacket displacements, indicating that the matter-wave dephasing is due to very weak potential
variation along the waveguide in our setup. We have demonstrated phase stability for an interferometry sequence with total interrogation time
exceeding one second. We also show that a four pulse interferometer can provide acceleration measurements with very long integration times that
are insensitive to apparatus vibrations, though it is important to note that the sensitivity of the interferometer scheme we describe
is compromised by the small wavepacket separations~\cite{Burke08}. 

In the future, such a system could study the quantum stability of wavepackets due to displaced potentials~\cite{phasespaceDispDiego} by
deliberately introducing time dependent variations in the potential along the waveguide direction ~\cite{kickedrotor95, stability08}. Instead of
measuring the mixed-state correlation functions, fidelity-type measurement~\cite{phasespaceDispDiego} can be proceeded with sub-recoil cooled
atoms occupying a single matter-wave state, where velocity-selective beamsplitting schemes can be applied~\cite{doublepulse05, Bragginf95}.

\begin{acknowledgments}
We thank helpful discussions from Prof. Eric Heller and Dr. Cyril
Petitjean. This work is supported in part by MURI and DARPA from
DOD, ONR and U.S. Department of the Army, Agreement Number
W911NF-04-1-0032, by NSF, and by the Charles Stark Draper
Laboratory.
\end{acknowledgments}

\end{document}